\documentclass{amsart}
\usepackage{amsfonts}

\usepackage{amscd}

\newtheorem{theorem}{Theorem}[section]

\newtheorem{proposition}[theorem]{Proposition}

\newtheorem{definition}{Definition}[section]

\theoremstyle{definition}
\theoremstyle{remark}
\numberwithin{equation}{section}

 2
 1

\begin{document}
\title[Geometrical and Operational Irreversibility]{Geometrical and Operational Aspects
of Irreversibility}
\author{P. Busch}
\address{Department of Mathematics, University of Hull, Hull HU6 7RX, UK}
\email{p.busch@maths.hull.ac.uk}
\date{October 7, 1999}
\dedicatory{Dedicated to E. Ruch on his 80th birthday.}
\maketitle

\begin{abstract}
In the statistical description of dynamical systems, an indication of the
irreversibility of a given state change is given geometrically by means of a
(pre-)ordering of state pairs. Reversible state changes of classical and
quantum systems are shown to be represented by isometric state
transformations. An operational distinction between reversible and
irreversible dynamics is given and related to the geometric characterisation
of the associated state transformations.
\end{abstract}

\section{Introduction}

In this paper a characterisation of the reversibility or irreversibility of
the time evolution of a dynamical system will be given that emphasises the
geometric structures underlying any statistical description.

The statistical description of a dynamical system is based on the dual
notions of states and observables. The states form a convex set of
probability measures (classical system) or density operators (quantum
system). These convex sets span in a natural way an ordered real vector
space, called \emph{state space}. For a classical system this is the space
of bounded (signed) measures on phase space; in the case of a quantum
statistical system, the state space is the set of self-adjoint trace class
operators over the system's Hilbert space. Observables are then represented
as bounded affine functionals on the set of states and hence as bounded
linear functionals on state space. This entails the description of
observables of a classical system as functions on phase space and of quantum
observables as self-adjoint operators on Hilbert space. In turn, a
statistical state can be viewed as a linear map on the real vector space of
observables, assigning to each observable its statistical average.

A convenient unified statistical description of classical as well as quantum
systems is thus given by the structure of a \emph{statistical duality} $%
\langle V,W\rangle $, where the state space $V$ is taken to be a complete
base norm space, with the convex base $K$ of the positive cone $V^{+}$
representing the set of states; and the space $W$ of observables is a
complete order unit space with closed order unit interval $E=[o,e]$ and such
that $W$ can be identified as a $\sigma (V^{*},V)$-dense subspace of $V^{*}$%
. The elements of $E$, called \emph{effects}, correspond to classes of
yes-no measurements that are indistinguishable in terms of their statistics.
The number $\langle x,f\rangle :=f(x)$ is interpreted as the probability for
registering a `yes' outcome in a measurement of the effect $f\in E$
performed on a system described by the state $x\in K$. For a lucid
introduction into the structure of a statistical duality, cf.\ Ref.\ \cite
{Stu88}. Recently the pair $\langle K,E\rangle $ has been the subject of
renewed interest and study and is commonly referred to as an instance of a 
\emph{statistical model}; the set of effects, $E$, is a realisation of an 
\emph{effect algebra} \cite{BelBug}.

The statistical description of the time evolution of a dynamical system is
based on the notion of a \emph{stochastic map} acting on a statistical state
space $V$, that is, a linear map $\Phi :V\to V$ that sends states to states, 
$\Phi (K)\subseteq K$. A stochastic map $\Phi $ is a contraction with
respect to the base norm: $||\Phi (z)||_1\le ||z||_1$ for all $z\in V$.
Hence the norm distance between two different states cannot increase under
the action of a stochastic map. This geometric property is taken up here to
formulate an operational characterisation for the (ir)reversibility of a
given stochastic map. A necessary condition for reversibility is that the
stochastic map under consideration must be an isometry. The converse
implication is put forward that irreversible dynamics are characterised, on
a suitable level of description, in terms of \emph{non-isometric} stochastic
maps. This conjecture is explored by means of some case studies of some
types of classical and quantum statistical systems. In contrast to the
conventional understanding, a reversible state transformation as defined
here is not necessarily surjective, though still always injective; but the
operational definition of reversible dynamics as a time-parameterised family
of reversible stochastic maps will be seen to force surjectivity.

The mathematical structure relevant to the subsequent investigation is
primarily that of a base norm space, while little use will be made of the
dual order unit space of observables. In a recent related work, a new way of
presenting the structure of a statistical state space has been developed
which emphasises the essential geometric and measure theoretic aspects of
this concept \cite{BuRu}. This reformulation is based on the concepts of 
\emph{measure cone} (representing the statistical state space), its
endomorphisms (which turn out to coincide with the stochastic maps) and, in
particular, the \emph{mixing distance}, an ordering relation of state pairs
that accounts for the dissimilarity of states. Previous investigations were
concerned with the fundamental geometric nature of these concepts \cite{Ru}
and their application in the context of statistical systems \cite{BuRu}.
Here the notion of mixing distance will be used to demonstrate the
connection between reversibility and isometric stochastic maps.

The present paper being based on \cite{BuRu}, notations, terminology and
basic facts are only briefly summarised.

\section{Statistical Description of Dynamical Systems}

\subsection{Statistical state space -- the measure cone}

The first definition describes the basic geometrical features of any
probabilistic framework.

\begin{definition}
A triple $(V,V^{+},e)$ is a measure cone if the following postulates are
satisfied:

(a) $V$ is a real vector space with convex, generating cone $V^{+}$ ($V\ =\
V^{+}-V^{+}$).

(b) $e:V\rightarrow \Bbb{R}$ is a linear functional, called \emph{charge},
that is strictly positive, 
\begin{equation}
z\in V^{+}\ \Longrightarrow \ \left\{ e(z)\geq 0,\ {\text{and}}\
e(z)=0\Leftrightarrow z=0\right\} .
\end{equation}
It follows that the charge $e$ admits a decomposition $e=e_{+}-e_{-}$ of $e$
into a difference of nonlinear, positive functionals $e_{\pm }$, where 
\begin{eqnarray}
\,\;\;e_{+} &:&V\to \Bbb{R}^{+},\ \ z\ \mapsto e_{+}(z):=\inf \left\{ e(x)|\
x\in V^{+},\ x-z\in V^{+}\right\} , \\
\,\;\;e_{-} &:&V\to \Bbb{R}^{+},\ \ z\ \mapsto e_{-}(z):=\inf \left\{ e(y)|\
y\in V^{+},\ z+y\in V^{+}\right\} .
\end{eqnarray}
Further, it is required that $e$ marks the cone contour: 
\begin{equation}
z\in V^{+}\ \ \Longleftrightarrow \ \ e(z)\ =\ e_{+}(z).  \label{co-cont}
\end{equation}

A measure cone $\left( V,V^{+},e\right) $\ is said to be a measure cone\
with \emph{minimal decomposition} if in addition the following postulate is
satisfied:

(c) To any $z\in V$ there exists a decomposition $z=z_{+}-z_{-}$, $z_{+},\
z_{-}\in V^{+}$ such that the following holds: $e(z_{+})=e_{+}(z),\
e(z_{-})=e_{-}(z)$. Any decomposition of $z$ with this property is called a
minimal decomposition of $z$.

A real vector space $V$ equipped with a measure cone\ $\left(
V,V^{+},e\right) $\ (with minimal decomposition) will be called mc-space
(with minimal decomposition).
\end{definition}

All known physically relevant examples of measure cones are equipped with a
minimal decomposition which is even unique. Hence in the sequel the term
measure cone shall generally be taken to include the existence of a minimal
decomposition.

The set $V^{+}$ is a proper (convex) cone so that $V$ becomes an ordered
vector space via $z\geq z^{\prime }\quad :\Leftrightarrow \quad z-z^{\prime
}\in V^{+}.$ The strict positivity of the charge functional $e$ ensures that
the intersection $K$ of the hyperplane $\{z\in V|e(z)=1\}$ with $V^{+}$ is a
base of the convex cone $V^{+}$. In a measure cone with minimal
decomposition the cone contour condition (\ref{co-cont}) is a consequence of
the strict positivity of $e$. Any vector space $V$ associated with a measure
cone can be equipped with a norm. More precisely, a triple $\left(
V,V^{+},e\right) $\ consisting of a real vector space $V$, a convex
generating cone $V^{+}\subset V$ and a linear functional $e$ is a measure
cone if and only if there exists a norm $\left\| \cdot \right\| $ marking
the cone contour in the following sense: 
\begin{equation}
z\in V^{+}\ \ \Longleftrightarrow \ \ e(z)={\Vert z\Vert }.
\end{equation}
In particular, the following is a norm of this type: 
\begin{equation}
{\Vert z\Vert }_1\ \ :=\ \ e_{+}(z)+e_{-}(z).
\end{equation}
This norm coincides with the the Minkowski functional of the set 
$B:=\text{co}(K\cup -K)$, 
the convex hull of $K\cup -K$, which makes $V$ a base norm
space (cf.~\cite{Alf}). The norm $||\cdot ||_1$ will be referred to as the
1-norm; it corresponds to the total variation norm in the classical case and
the trace norm in the quantum case.

It is worth noting that a decomposition $z=x-y$, $x,y\in V^+$, is a minimal
decomposition if and only if $\Vert x-y\Vert_1=\Vert x+y\Vert_1=e(x)+e(y)$.

The use of a measure theoretic terminology can be justified using the fact
that a base norm space $\left( V,\Vert \cdot \Vert _1\right) $ and its dual
order unit space $(V^{*},e)$ form a statistical duality. The set of effects
$E:=[o,e]\subset V^{*}$ is a partially ordered set of positive linear
functionals on $V$. $E$ is equipped with a complement operation, $a\mapsto
a^{\prime }:=e-a$ which induces a kind of weak orthogonality: effects $a,b$
are called \emph{orthogonal} if their sum $a+b$ is an effect again, that is,
if $b\leq a^{\prime }$. The elements $x$ of $V^{+}$ (of $K$), considered as
linear functionals on $V^{\prime }$ via $x(a):=a(x)$, act as positive,
additive [$x(a+b)=x(a)+x(b)$ whenever $a,b$ are orthogonal] (and normalised, 
$x(e)=1$) functions on $E$, representing thus (probability) measures in a
generalised sense.

\subsection{State transformations -- mc-endomorphisms\label{st-mc-end}}

The dynamics of a physical system is given by a family of state
transformations acting on its state space $K$. In agreement with the
statistical ensemble interpretation of the elements of $K$, a state
transformation should not alter the convex composition of a mixed state.
Hence a state transformation is an affine map; and as such it extends
uniquely to a linear map $\Phi :V\to V$ which is positive ($\Phi
(V^{+})\subset V^{+}$) and charge-preserving ($e\circ \Phi =e$). Such maps
will be referred to as \emph{mc-endomorphisms} of the mc-space with measure
cone $\left( V,V^{+},e\right) $\ generated by $K$ insofar as the geometric
aspect is concerned; bearing in mind the physical interpretation, the term 
\emph{stochastic map} will generally be used.

\begin{proposition}
\label{contr}Let $\left( V,V^{+},e\right) $\ be a measure cone\ equipped
with the 1-norm. \newline
(1) A stochastic map is a contraction.\newline
(2) A linear, charge-preserving contraction is positive, hence a stochastic
map.
\end{proposition}

\noindent 
\proof%
(1) Let $\Phi $ be a stochastic map. Then for $z\in V$, with minimal
decomposition $z=z_{+}-z_{-}$, 
\begin{equation*}
\left\| \Phi z\right\| _1\le \left\| \Phi z_{+}\right\| _1+\left\| \Phi
z_{-}\right\| _1=\left\| z_{+}\right\| _1+\left\| z_{-}\right\| _1=\left\|
z\right\| _1.
\end{equation*}
Hence $\Phi $ is a contraction.\newline
(2) Let $\Phi $ be linear, charge-preserving and contractive. Let $x\in
V^{+} $. Positive elements $z$ are characterised by thec cone contour
condition (\ref{co-cont}); hence we have to show that $\left\| \Phi
x\right\| _1=e(\Phi x)$. But we have $\left\| \Phi x\right\| _1\ge e(\Phi
x)=e(x)=\left\| x\right\| _1\ge \left\| \Phi x\right\| _1$, so that equality
must hold.%
\endproof%

The semigroup of stochastic maps induces a pre-ordering on the set of state
pairs $K\times K$: 
\begin{equation*}
(x,y)\sqsupseteq (x^{\prime },y^{\prime })\qquad :\Longleftrightarrow \qquad
(x^{\prime },y^{\prime })=(\Phi x,\Phi y)\quad \text{for some stochastic}\
\Phi .
\end{equation*}
In subsequent sections we will exhibit conditions under which the
sub-semigroup of stochastic isometries induces an equivalence relation on
$K\times K$, $(x^{\prime },y^{\prime })\equiv (x,y)$ iff $(x^{\prime
},y^{\prime })=(\Phi x,\Phi y)$ for some stochastic isometry $\Phi $. Hence
an equivalence class contains all state pairs that can be connected among
each other by means of some stochastic isometry. Then on the set of these
classes the above preordering becomes an ordering relation.

\subsection{Dissimilarity of states -- mixing distance}

The contractive nature of a stochastic map $\Phi $ leads to decreasing
distances (with respect to any mc-norm) between pairs of states from $K$
under the action of $\Phi $. More specifically, the action of $\Phi $ leads
to decreasing \emph{mixing distance}.

The \emph{mixing distance} of $x\in V^{+}\setminus \{0\}$ from $y\in
V^{+}\setminus \{0\}$ is defined as the map 
\begin{equation}
d[x/y]:\Bbb{R}^{+}\times \Bbb{R}^{+}\to \Bbb{R}^{+},\quad (\alpha ,\beta
)\mapsto \Vert \alpha x_0-\beta y_0\Vert _1
\end{equation}
($x_0:=x/\Vert x\Vert _1$, etc.). Two pairs $(x,y)$ and $x^{\prime
},y^{\prime })$ in $V^{+}\times V^{+}$ are called norm-equivalent if
$d[x/y]=d[x^{\prime }/y^{\prime }]$. Thus the mixing distance induces an
ordering on the classes of norm-equivalent pairs from $V^{+}\times V^{+}$
via 
\begin{equation}
d[x/y]\succ d[x^{\prime }/y^{\prime }]\ :\Longleftrightarrow \ \forall
\alpha ,\beta \in \Bbb{R}^{+}:\Vert \alpha x_0-\beta y_0\Vert _1\geq \Vert
\alpha x_0^{\prime }-\beta y_0^{\prime }\Vert _1.
\end{equation}

This concept is found to possess a canonical geometric interpretation in
terms of the direction distance, a norm-specific metric of angles in affine
spaces associated with a normed real vector space \cite{Ru}. The ensuing
ordering of angles suggests among others a notion of orthogonality which (in
the case of state pairs) corresponds to the idea of maximal mixing distance: 
$x,y\in K$ are called \emph{orthogonal}, $x\perp _1y$, if the following
condition is satisfied: 
\begin{equation}
\Vert \alpha x_0-\beta y_0\Vert _1\ \ =\ \ \Vert \alpha x_0+\beta y_0\Vert
_1\qquad \forall \alpha ,\beta \in \Bbb{R}^{+}.
\end{equation}
If $z=x-y$ is a minimal decomposition, then $x\perp _1y$; and conversely, if 
$x\perp _1y$ for $x,y\in V^{+}$, then $z=x-y$ is a minimal decomposition of
$z$ (\cite{BuRu}, Proposition 2.3).

\begin{proposition}
\label{stoch}Let $\left( V,V^{+},e\right) $ be a measure cone.\newline
(1) Any stochastic map $\Phi $ leads to decreasing mixing distance on
$K\times K$: 
\begin{equation*}
d[\Phi x/\Phi y]\ \prec \ d[x/y]\qquad \text{for}\qquad x,y\in K.
\end{equation*}
Hence, $(x,y)\sqsupseteq (x^{\prime },y^{\prime })\Rightarrow d[x/y]\succ
d[x^{\prime }/y^{\prime }]$.\newline
(2) A stochastic map $\Phi $ is an isometry (hence preserving the mixing
distance) if and only if $\Phi $ is orthogonality-preserving: 
\begin{eqnarray*}
\Vert \Phi z\Vert _1=\Vert z\Vert _1\qquad \forall z\in V\qquad
&\Longleftrightarrow &\qquad d[\Phi x/\Phi y]=d[x/y]\qquad \forall x,y\in K
\\
&\Longleftrightarrow &\qquad \left( x\perp _1y\Rightarrow \Phi x\perp _1\Phi
y\right) \forall x,y\in K.
\end{eqnarray*}
\end{proposition}

\noindent 
\proof%
(1) This is an immediate consequence of the fact that $\Phi $ is a
contraction.\newline
(2) Let $\Phi $ be an orthogonality-preserving stochastic map. Then 
\begin{eqnarray*}
\Vert \Phi z\Vert _1 &=&e(\Phi z_{+})+e(\Phi z_{-})\qquad [\text{$\Phi $
positive, orthogonality-preserving}] \\
&=&e(z_{+})+e(z_{-})\qquad [\text{$\Phi $ charge-preserving}] \\
&=&\left\| z\right\| _1.
\end{eqnarray*}
Conversely, assume $\Phi $ to be a stochastic isometry; then for the minimal
decomposition of $z\in V$, $z=z_{+}-z_{-}$ one has 
\begin{equation*}
e(\Phi z_{+})+e(\Phi z_{-})=e(z_{+})+e(z_{-})=\left\| z_{+}-z_{-}\right\|
_1=\left\| \Phi z_{+}-\Phi z_{-}\right\| _1,
\end{equation*}
so that $\Phi z=\Phi (z_{+})-\Phi (z_{-})$ is a minimal decomposition as
well and therefore orthogonal. Thus if $z=x-y$ for any orthogonal pair
$x,y\in V^{+}$, then $\Phi x,\Phi y$ is an orthogonal pair.
\endproof

Statement (1) describes the crucial role of the mixing distance as an
indicator of irreversibility: if the mixing distance decreases under a
stochastic map $\Phi $, then $\Phi $ cannot be an isometry, so that there is
no stochastic map that would reverse the action of $\Phi $. In this sense
the mixing distance has the same function as the (relative) entropy.
However, it is known that under certain conditions (though not in general)
the converse of statement to (1) holds, thus showing that the mixing
distance is superior to entropy insofar as its decrease between two state
pairs is even sufficient to ensure the existence of: a state transformation
that connects the pair.

\begin{theorem}
\label{rss}Let the measure cone $\left( V,V^{+},e\right) $ be given by
$V=L^1(\Omega ,\Sigma ,\mu )$, with $(\Omega ,\Sigma ,\mu )$\ a separable,
$\sigma $-finite measure space. The following is true: given two pairs of
states $x,y$ and $x^{\prime },y^{\prime }$ such that $d[x^{\prime
}/y^{\prime }]\prec d[x/y]$, then there exists a stochastic map which
transforms $x$ into $x^{\prime }$ and $y$ into $y^{\prime }$. That is: 
\begin{equation*}
(x,y)\sqsupseteq (x^{\prime },y^{\prime })\Leftrightarrow d[x/y]\succ
d[x^{\prime }/y^{\prime }].
\end{equation*}
\end{theorem}

\noindent In this form the theorem has been proved in \cite{BuQu}. The
theorem was initially found in a more specific form as a generalisation of a
theorem due to Hardy, Littlewood and Polya on doubly stochastic matrices\cite
{RSS}. Recently this result has been exhaustively generalised by Ruch and
Stulpe to cover all conceivable ``classical'' spaces of measures 
\cite{RuStu}.

\section{Irreversibility}

The convex semigroup of stochastic maps acts transitively on the set $K$.
Hence any transition $x\to x^{\prime }$ is physically realisable in the
sense that there exists a stochastic map $\Phi $ such that $x^{\prime }=\Phi
x$. As a consequence, the phenomenon of irreversibility can manifest itself
only if at least pairs of states and their transitions are taken into
consideration \cite{Ru}. According to Proposition \ref{stoch}, decreasing
mixing distance is a necessary condition for the possibility of transforming 
$x,y$ into $x^{\prime },y^{\prime }$ by means of one and the same stochastic
map. Thus, if $d[\Phi x/\Phi y]\neq d[x/y]$, then there is no stochastic map
transforming both $x^{\prime },y^{\prime }$ back into $x,y$. In this sense,
strict decrease of the mixing distance between two state pairs is an
indicator of the irreversibility of the stochastic map $\Phi $.

\begin{definition}
\label{irr}A stochastic map $\Phi $ acting on an mc-space is irreversible if
and only if there is a pair $(x,y)\in K\times K$ such that $(\Phi x,\Phi y)$
cannot be transformed back into $(x,y)$, i.e., $(\Phi x,\Phi y)\not%
{\sqsupseteq}(x,y)$.
\end{definition}

\noindent In general a physical ``reversal of motion'' involves a
time-inversion operation $\Theta $, represented as a positive surjective
isometry on $V$. Thus irreversibility as defined above is equivalent to
$(\Theta \Phi x,\Theta \Phi y)\not{\sqsupseteq}(\Theta x,\Theta y)$, as it
should.

An immediate consequence of Proposition \ref{contr} is the following.

\begin{proposition}
A reversible stochastic map on an mc-space is an isometry.
\end{proposition}

\noindent In cases where Theorem \ref{rss} is valid it follows that a
stochastic map $\Phi $ is reversible whenever for arbitrary state pairs
$x,y\in K$ one has $d[x/y]\prec d[\Phi x/\Phi y]$. But this amounts to saying
that $\Phi $ is an isometry. Hence, one has the following result.

\begin{theorem}
\label{revers}Let $V$ be an mc-space equipped with 1-norm in which the
statement of Theorem \ref{rss} holds. Then a stochastic map $\Phi $ on $V$
is reversible if and only if it is an isometry. This is the case exactly
when the mixing distance is invariant under $\Phi $.
\end{theorem}

\noindent Within the domain of validity of this theorem, the symmetry of the
relation $\left( x^{\prime },y^{\prime }\right) \equiv \left( x,y\right) $
defined at the end of subsection \ref{st-mc-end} is thus established, so
that this relation is an equivalence relation.

Definition \ref{irr} constitutes what we referred to as an operational
characterisation of the irreversibility of state changes. Theorems \ref{rss}
and \ref{revers} provide the foundation for the geometric indication of
irreversibility via strictly decreasing mixing distance. Theorem \ref{rss}
also gives a sufficient criterion for the operational realisability
(existence of a stochastic map) of certain changes (jointly sending state
pair $x,y$ to state pair $x^{\prime },y^{\prime }$). The question arises
whether strictly dereasing mixing distance under the action a stochastic map 
$\Phi $, or equivalently, lack of the isometric property of $\Phi $, fully
captures the physical content of the notion of irreversibility. The study of
irreversibility is primarily concerned with dynamical processes taking place
over a period of time rather than for a single time step. Thus reversibility
or irreversibility is to be regarded as a property of a (statistical)
dynamical system, represented as a time-parameterised family of stochastic
maps, $\left( \Phi _t\right) _{t\in T}$, with $T=[0,\infty )$ or $T=\Bbb{R}$
. (For simplicity we assume homogeneity of time and allow for time to extend
to the infinite future (and past); hence for any time $t_0$, the transition
to $t_0+t$ is given by $\Phi _t$). Moreover, it is important to bear in mind
that the irreversible behaviour of a dynamical system emerges on a certain
level of description, usually referred to as macroscopic or thermodynamic.
This has led to the well-known problem of reconciling the omnipresence of a
time arrow in large-scale phenomena with the microscopic description of the
world which is usually taken to be based on the fundamentally reversible
dynamical laws (of Newtonian mechanics or quantum mechanics). Without going
into further detail, we recall that the statistical description ((quantum)
statistical mechanics) was introduced as a basis for any attempt to
formulate a consistent bridge between the two (microscopic and macroscopic)
levels of description. In fact, statistical models as defined in Section 2
can be viewed as a convenient unified framework for formalising all kinds of
coarse-grainings used to reflect the coarseness of macroscopic observations
as well as the tracing out of unobservable degrees of freedom.

The fact that in the modelling of real physical systems there is usually a
hierarchy of levels of descriptions shows that a characterisation of the
irreversibility or reversibility of the observed dynamics must depend on the
level of description appropriate to the feasible observations. Thus the
formal definitions of (ir)reversibility for (a) a single stochastic map and
(b) for a statistical dynamical system $\left( \Phi _t\right) _{t\in T}$ are
not in themselves sufficient to characterise the irreversible behaviour of a
physical system but they must be supplemented with a specification of the
appropriate level of description to which they pertain. We believe the
following definition captures those features that are commonly accepted as
characteristic of irreversible physical processes. To formulate the notion
of a reversed process, one must postulate the existence of a stochastic map
$\Theta $ which represents the action of time inversion, or more precisely,
motion reversal. As a double application of $\Theta $ should restore the
original state of motion, $\Theta $ must equal its own inverse and thus is a
bijective stochastic isometry.

\begin{definition}
\label{rev-dyn} A dynamical system $\left( \Phi _t\right) _{t\in T}$ on a
statistical state space $V$, with time inversion operation $\Theta $, is 
\emph{reversible} if for all $t$, 
\begin{equation}
\Theta ^{-1}\Phi _t\Theta \mid _{\Phi (V)}=\Phi _t^{-1}.  \label{t-rev}
\end{equation}
\end{definition}

\noindent Note that this concept does not stipulate the state transformation
to be surjective. It thus represents exactly the idea of reversing a given
state change, by means of the \emph{same }dynamics, without changes in the
environment. Usually the condition of time reflection symmetry, $\Phi
_t^{-1}=\Phi _{-t}$, is taken to be part of the concept of reversibility, so
that a semigroup $\left( \Phi _t\right) _{t\in T}$ actually would have to
extend to a group in order to be reversible.

These considerations will be illustrated with a number of case studies.

\section{Case Studies}

\subsection{Classical Dynamical Systems}

In this section $\left( \Omega ,\Sigma ,\mu \right) $ denotes a separable,
$\sigma $-finite measure space and $V=V_c$ the ``classical'' mc-space
corresponding to the real-valued, bounded, $\sigma $-additive set functions
on $\left( \Omega ,\Sigma \right) $ which are absolutely continuous with
respect to $\mu $. Hence, $V_c$ is isomorphic to the Banach space $%
L^1(\Omega ,\Sigma ,\mu )$. In this situation Theorem \ref{rss} can be used
to obtain a characterisation of reversible state transformations which is
based on the existence of an inverse map. In general it need not be true
that the inverse of a stochastic map $\Phi $ (if it exists) can be extended
from the range of $\Phi $ to all of $V_c$, nor that it is positive itself; a
non-positive inverse, or one that cannot be extended, does not have a
physical interpretation as a state transformation.

\begin{proposition}
\label{strev}A stochastic map $\Phi :V_c\to \Phi (V_c)$ is reversible if and
only if there exists an inverse map $\Phi ^{-1}:{\Phi }(V_c)\to V_c$ which
is positive.
\end{proposition}

\proof 
Let $\Phi $ be reversible and therefore, by Theorem \ref{revers}, an
isometry. The range ${\Phi }(V_c)$ is a closed subspace of $V_c$, thus a
base norm space itself. Due to the injectivity of $\Phi $ the inverse $\Phi
^{-1}$ exists on ${\Phi }(V_c)$ and is a charge-preserving contraction (in
fact, an isometry). $\Phi ^{-1}$ is positive; otherwise there were an
element $z$ with minimal decomposition $z=z_{+}-z_{-}$, $z_{\pm }\in
V_c^{+}\setminus \{0\}$ such that $\Phi z\in V_c^{+}$, in contradiction to
the fact that $\Phi z_{-}\neq 0$ (note that $\Phi $ is
orthogonality-preserving). \newline
Conversely, if the inverse $\Phi ^{-1}:{\Phi }(V_c)\to V_c$ exists and is
positive (it is automatically charge-preserving), then $\Phi $ is
necessarily an isometry, hence, by Theorem \ref{revers}, reversible.
\endproof%

As noted above, the reversibility of a dynamical system is sometimes defined
by means of the group property of the respective family of state
transformations. We have introduced a general (statistical) dynamical system
as a semigroup $\left( \Phi _t\right) _{t\geq 0}$ of stochastic maps acting
on $V_c$. More specifically, given a measure space $\left( \Omega ,\Sigma
,\mu \right) $, we will now consider dynamical systems defined as a
semigroup $\left( S_t\right) _{t\geq 0}$ of measurable maps $S_t:\Omega \to
\Omega $ which leave $\mu $ invariant. Then the associated semigroup of
stochastic maps is determined via 
\begin{equation*}
\int_\Delta \Phi _t\rho d\mu \ :=\ \int_{S_t^{-1}(\Delta )}\rho d\mu ,\qquad
\rho \in V_c,\ \Delta \in \Sigma .
\end{equation*}
In the case of a normalised measure space ($\mu (\Omega )=1$), the uniform
distribution $\rho _u=1_\Omega $ is a fixed point of all $\Phi _t,\ t\geq 0$.
In accordance with Definition \ref{irr} a semigroup $\left( \Phi _t\right)
_{t\geq 0}$ of stochastic maps shall be called reversible if all $\Phi _t$
are reversible. While the group property is sufficient to ensure
reversibility in the sense of Definition \ref{irr}, it is not in general a
necessary condition as will be shown by means of an example below. However,
for a fairly general class of dynamical systems the group property is
necessary and sufficient for reversibility. The following result is due to
R. Quadt and the author and was originally published in \cite{Qu}.

\begin{proposition}
Let $\left( \Omega ,{\mathcal{B}}(\Omega ),\mu \right) $ be a normalised
measure space, with $\left( \Omega ,{\mathcal{B}}(\Omega )\right) $ a
standard Borel space. Let $\left( S_t\right) _{t\geq 0}$ be a dynamical
system, with induced semigroup $\left( \Phi _t\right) _{t\geq 0}$ and time
inversion operation $\Theta $. $\left( \Phi _t\right) _{t\geq 0}$ of
stochastic maps is reversible if and only if it can be extended to a group
via $\Theta ^{-1}\Phi _t\Theta =:\Phi _{-t}$.
\end{proposition}

\proof
That the group property is sufficient for reversibility is clear from
Proposition \ref{strev}. Conversely, let $\left( \Phi _t\right) _{t\geq 0}$
be reversible. By Theorem \ref{revers} all the $\Phi _t$ are isometries. To
ensure the group extension, one shows that the $\Phi _t$ are surjective. To
this end one constructs a family $\gamma _t:\Omega \to \Omega $ of
measurable, $\mu $-preserving, surjective point maps such that $\Phi _t\rho
(x)=\rho \circ \gamma _t(x)$ ($\mu $-almost everywhere) for $\rho \in V_c$.
The maps $\gamma _t$ will turn out to be uniquely determined up to Borel
sets of measure zero. It then follows that an inverse map $\Phi _t^{-1}$ is
defined on all of $V_c$ via $\int_\Delta \Phi _t^{-1}\rho d\mu
:=\int_{\gamma _t^{-1}(\Delta )}\rho d\mu $. Then $\Phi _{-t}=\Phi _t^{-1}$,
and the group property is established. To find $\gamma _t$, note that since
$\left( \Omega ,{\mathcal{B}}(\Omega )\right) $, is a standard Borel space,
there exists a measurable, bijective map $\psi :\Omega \to [0,1]$ which
induces a bijective isometry $j:V_c\to L^1\left( [0,1],{\mathcal{B}}
([0,1]),\nu \right) $ via $\int_{\widetilde{\Delta }}j\rho d\nu :=\int_{\psi
^{-1}(\widetilde{\Delta })}\rho d\mu $ and $\nu (\widetilde{\Delta }):=\mu
\left( \psi ^{-1}(\widetilde{\Delta })\right) $. Here $\left( [0,1],{
\mathcal{B}}([0,1]),\nu \right) $ is a normalised, separable measure space.
It follows that the map $\widetilde{\Phi }_t:=j\circ \Phi _t\circ j^{-1}$ is
an isometry on $L^1\left( [0,1],{\mathcal{B}}([0,1]),\nu \right) $. By
Lamperti's theorem \cite{Roy}, there exists a measurable, surjective map
$\varphi _t:[0,1]\to [0,1]$ (unique up to Borel sets of $\nu -$measure zero)
such that $\widetilde{\Phi }_tf=\widetilde{\Phi }_t1_{[0,1]}\cdot f\circ
\varphi _t$ and $\int_{\varphi _t^{-1}(\widetilde{\Delta })}
\widetilde{\Phi }_t1_{[0,1]}d\nu =\int_{\widetilde{\Delta }}d\nu $. 
Since $\widetilde{\Phi }_t1_{[0,1]}=1_{[0,1]}$, 
it follows that $\varphi _t$ is measure-preserving.
Now, using the equation $j\rho (x)=\rho \circ \psi ^{-1}(x)$ (valid almost
everywhere), one obtains the desired result: $\Phi _t\rho (x)=\left(
j^{-1}\circ \widetilde{\Phi }_t\circ j\rho \right) (x)=\rho \circ \psi
^{-1}\circ \varphi _t\circ \psi (x)=:\rho \circ \gamma _t(x)$ (valid almost
everywhere).
\endproof

There exist semigroups of reversible stochastic maps which do not admit an
extension to a group. As an example, let 
$\left( \Bbb{R},\mathcal{B}(\Bbb{R}),\mu _L\right) $ 
be the Borel-Lebesgue measure space. It is easy to
construct a measurable bijection $\gamma :\Bbb{R}\to (0,\infty )$ which,
together with its inverse $\gamma ^{-1}:(0,\infty )\to \Bbb{R}$, is
measure-preserving. For example, consider a partitioning of the real line
into intervals of the form $(n,n+1]$, $n$ integer. If the label $n$ is even
(odd), call the corresponding interval even (odd). Now the map $\gamma $ may
be defined by shifting the positive (negative) intervals one by one with
increasing $|n|$ onto the positive even (odd) intervals with correspondingly
increasing labels. With $\gamma _n:=\gamma ^n$ one defines a (discrete)
semigroup of transformations on $\Bbb{R}$ such that the induced family of
linear operators $\left( \Phi _n\right) ,\ \Phi _n\rho :=\rho \circ \gamma
_n $ on $V_c$ is a semigroup of isometric stochastic maps. By Theorem \ref
{revers} the stochastic maps $\Phi _n$ are reversible; but $\left( \Phi
_n\right) _{n\in \Bbb{N}_0}$ does not have an extension to a group since the 
$\Phi _n^{-1}$ cannot be extended to isometries on $V_c$. So if one could
construct a bijective stochastic isometry $\Theta $ such that $\Theta
^{-1}\Phi _1\Theta =\Phi _1^{-1}$, one would have found an example of a
reversible dynamical semigroup which does \emph{not} admit a group
extension. The crucial point of this example is that the underlying measure
space is not finite, so that proper subsets of $\Omega $ are measure
theoretically equivalent to $\Omega $ itself. Redistribution operations such
as $\gamma $ can be applied, for instance, as a coding of the set $\Bbb{R}$
into $(0,\infty )$.

\subsection{Damped Motion}

As an example of a deterministic dynamical system that is not measure
preserving we consider the simple case of linearly damped motion of a
particle in one dimension. Thus the state of the particle at any time $t$ is
given by its position $X\left( t\right) $ and velocity $\dot{X}\left(
t\right) $, that is, $\omega =\left( X,\dot{X}\right) \in 
\Omega =\Bbb{R}^2$. 
The dynamics is determined by the equation of motion $\ddot{X}=-\kappa 
\dot{X}$, $\kappa >0$, which is solved by 
\begin{equation*}
S_t:\left( X\left( 0\right) ,\dot{X}\left( 0\right) \right) \mapsto \left(
X\left( t\right) ,\dot{X}\left( t\right) \right) =\left( X\left( 0\right) +
{\textstyle{\frac 1\kappa}} \dot{X}\left( 0\right) \left( 1-e^{-\kappa
t}\right) \,,\,\dot{X}\left( 0\right) e^{-\kappa t}\right) .
\end{equation*}
It is easy to verify that $S_t^{-1}=S_{-t}$, so that $\left( S_t\right)
_{t\in \Bbb{R}}$ is a group. But the latter maps, $S_{-t}$, are seen to
solve the anti-damping equation $\ddot{X}=+\kappa \dot{X}$, which is
obtained from the previous one by application of the time inversion map
$\theta :\left( X,\dot{X}\right) \mapsto \left( X,-\dot{X}\right) $.
Accordingly, we find that $\theta ^{-1}S_t\theta \ne S_t^{-1}$, which
carries over in the corresponding inequality $\Theta ^{-1}\Phi _t\Theta \ne
\Phi _t^{-1}$ for the induced stochastic semigroup, with all $\Phi _t$
surjective stochastic isometries on $L^1\left( \Omega ,\mathcal{B}\left(
\Omega \right) \right) $. This confirms that the damped motion is
irreversible, despite the fact that a formal extension to a group is
possible. A natural indicator of the irreversibility (Lyapounov variable) is
given by the magnitude of the velocity, $\left| \dot{X}(t)\right| =\left| 
\dot{X}(0)\right| e^{-\kappa t}$, which tends monotonically to 0 as $t$
increases.

Damped motion of a particle can be viewed as a reduced description of a
system consisting of a very massive body suspended in a medium (gas or
fluid) of molecules with which it interacts via collisions. Despite the
presence of the environment, the body performs a deterministic motion
whereas its energy is dissipated into the degrees of freedom represented by
the molecules of the medium (as well as increase of internal heat of the
body). The next example of Brownian motion belongs to the same physical
class but the body suspended in the medium is not as massive so that its
motion is randomised due to unobservable collisions with the surrounding
molecules.

\subsection{Brownian Motion}

The random collisions determining the motion of a Brownian particle are
modelled by means of a stochastic differential equation for its position,
$X\left( t\right) $: 
\begin{equation*}
\dot{X}=b(X)+\sigma (X)\xi .
\end{equation*}
Here $b\left( X\right) $ describes a deterministic influence while the white
noise term $\xi =\dot{w}$ is given as the time derivative of a Wiener
process; $\sigma (X)$ is the amplitude of the stochastic perturbation. As is
well known, this stochastic process can be represented in terms of an
associated Fokker-Planck (or Kolmogorov) equation for density functions
$\rho _t(X)$, 
\begin{equation*}
\frac{\partial \rho _t}{\partial t}=-\frac{\partial \left[ b(X)\rho
_t\right] }{\partial X}+\frac 12\frac{\partial ^2\left[ \sigma ^2\left(
X\right) \rho _t\right] }{\partial X^2},
\end{equation*}
the solution of which (for sufficiently regular amplitude $\sigma (X)$) is
given by an \emph{exact} semigroup $\left( \Phi _t\right) _{t\ge 0}$;
exactness meaning that $\Phi _t\rho $ converges in 1-norm to a unique
stationary distribution $\rho ^{*}$ \cite{Mack}. This process is thus
characterised by decreasing mixing distance between any density $\rho _t$
and $\rho ^{*}$, in agreement with the fact that there exist Lyapounov
variables indicating the irreversibility.

\subsection{Instability}

The preceding examples display irreversible behaviour of a system due to its
interaction with a (stationary) environment. An alternative type of
situation is given by \emph{closed} deterministic systems which are
characterised by a degree of intrinsic \emph{instability}. Thus it is known
that for the so-called \emph{K-systems} there are dynamics-dependent coarse
grainings under which the \emph{observable} motion is described by a
semigroup of strictly contractive stochastic maps (e.g., \cite{GMC}).
Alternatively, a dynamical system $\left( S_t\right) _{t\in \Bbb{R}}$ is
called \emph{intrinsically random }if its associated group of stochastic
isometries $\left( \Phi _t\right) _{t\in \Bbb{R}}$ is similar to a semigroup
of strictly contractive stochastic maps $\left( \widetilde{\Phi }_t\right)
_{t\ge 0}$; this means that there is an invertible stochastic map $W$ whose
inverse has dense domain and is \emph{not }positive such that $\widetilde{
\Phi }_t=W\Phi _tW^{-1}$. It has been shown that K-systems possess this
property of intrinsic randomness and that for them the irreversibility of
the stochastic semigroup $\left( \widetilde{\Phi }_t\right) _{t\ge 0}$ can
indicated by some Lyapounov variables \cite{GMC,MPC}.

\subsection{Quantum Mechanics}

One may consider the conjecture that the assertion made in Theorem \ref
{revers} remains true even beyond the scope of Theorem \ref{rss}. This
question shall be investigated in the context of quantum mechanical measure
cones for which Theorem \ref{rss} is known to be violated unless the
underlying Hilbert space is two-dimensional \cite{AlUhl}, see also the
corresponding remarks in \cite{BuRu}. One can construct quantum mechanical
stochastic maps that are isometric and reversible without being surjective
but such that their inverse maps can be extended to stochastic maps.

Let $\mathcal{H}$ denote a separable complex Hilbert space (with inner
product $\langle\cdot|\cdot\rangle$ associated to a quantum mechanical
system. The ensuing mc-space $V=V_q$ is given by the Banach space of
selfadjoint trace class operators, with $K=K_q$ representing the set of
density operators. The charge functional and 1-norm are given by the trace
and trace norm, respectively. The surjective isometries among the stochastic
operators possess a particularly simple structure.

\begin{proposition}
Let $V_q$ be the mc-space associated with a separable complex Hilbert space
$\mathcal{H}$. A surjective stochastic map $\Phi :V_q\to V_q$ is an isometry
if and only if it is induced by a linear or antilinear isometry 
$U:\mathcal{H}\to \mathcal{H}$ such $\Phi (z)=UzU^{*}$ for $z\in V_q$.
\end{proposition}

This fact follows readily from the Wigner--Kadison characterisation of the
automorphisms of states or observables \cite{Bar,Kad}. We present a concise
proof that makes use of a result of Davies \cite{Dav}.

\proof
First, any $\Phi $ defined as above in terms of some unitary or antiunitary
$U$ is a positive, trace-preserving map on $V_q$. This follows from the fact
that $U^{*}U=I$: $e(UzU^{*})=e(U^{*}Uz)=e(z)$, the first equality being due
to the invariance of the trace under cyclic permutations of the factors in
its argument. Let $z\in V_q^{+}$, then $\langle \varphi |UzU^{*}\varphi
\rangle =\langle U^{*}\varphi |zU^{*}\varphi \rangle \geq 0$ for all
$\varphi \in \mathcal{H}$; hence, $\Phi (z)$ is positive, too. To verify the
isometric nature of $\Phi $, let $z=z_{+}-z_{-}$ be a minimal decomposition.
It follows that $z_{+}\cdot z_{-}=0$ and therefore $Uz_{+}U^{*}\cdot
Uz_{-}U^{*}=U(z_{+}\cdot z_{-})U^{*}=0$. Thus, $\Phi (z_{+})$ and $\Phi
(z_{-})$ are orthogonal so that $\Phi (z)=\Phi (z_{+})-\Phi (z_{-})$ is a
minimal decomposition. Since $\Phi $ is trace-preserving it follows that
$\Vert \Phi (z)\Vert _1=e(z_{+})+e(z_{-})=\Vert z\Vert _1$.\newline
Next, let $\Phi $ be a surjective isometric stochastic map. Then it is also
It follows that $\Phi $ is a \textsl{pure} map sending pure (extremal)
states to pure states: indeed, assume $x\in K_q$ is pure, let $\Phi
(x)=\lambda y_1+(1-\lambda )y_2$ for some $y_1,y_2\in K_q$ and 
$0<\lambda <1$. 
Since $\Phi $ is surjective there exist $x_1,x_2\in K_q$ such that $\Phi
(x_1)=y_1$, $\Phi (x_2)=y_2$. By the injectivity of $\Phi $, $x=\lambda
x_1+(1-\lambda )x_2$, and due to the purity of $x$, $x_1=x_2=x$; therefore
$y_1=y_2=y$, that is, $y=\Phi (x)$ is pure, too. According to Theorem 2.3.1
of \cite{Dav}, $\Phi $ is induced by a unitary or antiunitary operator.
\endproof

If in the case of an infinite-dimensional Hilbert space the assumption of
surjectivity is dropped, then there exists a class of non-pure stochastic
isometries which can be constructed as follows.

\begin{proposition}
Let $\mathcal{H}=\mathcal{H}_0\oplus \mathcal{H}_1\oplus {}_2\oplus \cdots
\oplus \mathcal{H}_n$ be a direct sum decomposition of $\mathcal{H}$ such
that $\text{dim}\mathcal{H}_k=\infty $, $k=1,2,\cdots ,n$, $2\leq n\leq
\infty $. Let $U_k:\mathcal{H}\to \mathcal{H}_k$ be linear or antilinear
isometries, $0\leq w_k\leq 1$, $\sum w_k=1$. Then 
\begin{equation}
\Phi :V_q\to V_q,\ \ z\ \mapsto \ \Phi (z)\ :=\ \sum w_kU_kzU_k^{*},
\label{st-is}
\end{equation}
is an isometric stochastic map. Moreover, the following is a stochastic map
whose restriction to the range of $\Phi $ coincides with the inverse of
$\Phi $: Let $P_0,\ P_k=U_kU_k^{*}$ denote the orthogonal projections
associated to the subspaces $\mathcal{H}_0,\ \mathcal{H}_k$, respectively. 
\begin{equation}
\Psi :V_q\to V_q,\ \ z\ \mapsto \ \Psi (z)\ :=\
\sum_{k=1}^nU_k^{*}P_kzP_kU_k\ +\ P_0zP_0.  \label{invers}
\end{equation}
\end{proposition}

\proof
It is obvious that $\Phi $ is a stochastic map. The isometric nature follows
from the fact that all the $U_k(z_{\pm })$, $U_l(z_{\pm })$ (for minimal
decompositions $z=z_{+}-z_{-}$ and $k\neq l$) are mutually orthogonal, so
that $\Vert \sum w_kU_kzU_k^{*}\Vert _1=\sum w_k\Vert U_kzU_k^{*}\Vert
_1=\sum w_k\Vert z\Vert _1$.

The positivity of $\Psi $ is obvious. It follows from $\sum_{k=0}^nP_k=I$
that $\Psi $ is trace-preserving. Finally, for any element $\Phi (z)$ one
has $P_k\Phi (z)P_k=w_kU_kzU_k^{*}$ and $P_0\Phi (z)P_0=0$. This immediately
yields $\Psi \left( \Phi (z)\right) =\sum w_kz=z$.%
\endproof%

The last result shows that isometric state transformations of the form (\ref
{st-is}) are indeed reversible. It is known that all isometric stochastic
maps on $V_q$ are of this form \cite{Bu99}.

While the statement of Theorem \ref{rss} does not in general hold in quantum
mechanics, the last result entails that for pairs of quantum states, the
relation $\left( x^{\prime },y^{\prime }\right) \equiv \left( x,y\right) $
(subsection \ref{st-mc-end}) is again symmetric. Hence it is an equivalence
relation and renders the relation $\sqsupseteq $ a partial ordering on the
ensuing equivalence classes. These classes contain as a subclass those state
pairs that can be connected with a \emph{surjective} stochastic isometry. In
the above quantum mechanical example it becomes apparent that this specific
subclass is strictly smaller than the original equivalence class. In fact,
the surjective stochastic isometries are those induced by either unitary or
antiunitary maps and hence always send pure states to pure states. By
contrast, the map (\ref{st-is}) sends pure states to mixed states whenever
it is not surjective. Thus a pair of image states cannot be sent to a pair
of pure states by means of a surjective stochastic isometry. The implication
of this observation is that non-surjective maps of the form (\ref{st-is})
cannot be interpreted as (discrete-time) reversible dynamics: the maps (\ref
{invers}) are not stochastic isometries themselves, so that requirement (\ref
{t-rev}) of Definition \ref{rev-dyn} cannot be satisfied for a statistical
dynamical system consisting solely of stochastic isometries.

\section{Conclusion}

In this work we have reviewed the operational characterisation of
irreversible dynamical processes and have explored the possibility of an
intrinisically geometrical indication of reversibility or irreversibility,
based on the fundamental concept of mixing distance introduced by E. Ruch.
We have reviewed this concept in the abstract language of statistical
dualities which provides a unified framework for classical and quantum
statistical theories an moreover brings out the essential geometric features.

Irreversibility of a \emph{single} statistical state transformation is
defined as the impossibility of undoing the change of some pairs of states
by application of another state transformation. It follows that a reversible
stochastic map is necessarily an isometry. On the other hand, stochastic
isometries which are surjective are reversible. The conjecture is proposed
that all stochastic isometries are reversible. On the basis of the principle
of decreasing mixing distance (Theorem \ref{rss}) this conjecture is
verified for certain classical cases. An explicit classification of quantum
mechanical stochastic isometries yields the same result for quantum
statistical systems. Hence reversible state transformations are necessarily
isometric, that is, they leave the mixing distance for state pairs
invariant; but they are \emph{not} necessarily surjective.

The full physical content of the notion of reversibility cannot solely be
represented as a metric property involving the mixing distance; in addition
one needs to make explicit the notion of motion reversal, which involves a
bijective isometric stochastic time inversion map $\Theta $. Physically,
reversibility means that it is the \emph{same }dynamical map $\Phi _t$ that
leads back to the initial state if applied to the motion-reverted final
state: 
\begin{equation*}
x\rightarrow \Phi _tx\rightarrow \Theta \Phi _tx\rightarrow \Phi _t\Theta
\Phi _tx\rightarrow \Theta ^{-1}\Phi _t\Theta \Phi _tx=x.
\end{equation*}
This again entails that the inverse to $\Phi _t$ is positive and
charge-preserving on its domain and hence a stochastic map; thus any
reversible statistical dynamical system $\left( \Phi _t\right) $ must be
composed of isometric stochastic maps. The possibility remains that
reversible dynamics may not in every case be given by surjective stochastic
isometries. To summarise, invariance of mixing distance is necessary for
reversibility and decrease of mixing distance is an indication of
irreversibility. The power of the concept of mixing distance in the context
of classical statistical systems lies in the fact that its decrease provides
a sufficient criterion for the physical realisability of joint changes of
state pairs as expressed in Theorem \ref{rss}.

\end{document}